\documentclass[preprint,aps,showkeys,showpacs]{revtex4}

\usepackage{amsmath}

\begin{document}

\title{Lorentz Transformation of Blackbody Radiation.}
\author{G. W. Ford}
\affiliation{Department of Physics, University of Michigan, Ann Arbor, MI 48109-1040 USA}
\author{R. F. O'Connell\thanks{
oconnell@phys.lsu.edu}}
\affiliation{Department of Physics and Astronomy, Louisiana State University, Baton
Rouge, LA 70803-4001 USA}
\date{\today }

\begin{abstract}
We present a simple calculation of the Lorentz transformation of the
spectral distribution of blackbody radiation at temperature $T$. Here we
emphasize that $T$ is the temperature in the blackbody rest frame and does
not change. We thus avoid the confused and confusing question of how
temperature transforms. We show by explicit calculation that at zero
temperature the spectral distribution is invariant. At finite temperature we
find the well known result familiar in discussions of the the 2.7${{}^\circ}$
K cosmic radiation.
\end{abstract}

\maketitle

Here we present a straightforward derivation of the spectral distribution of
blackbody radiation as seen in a moving frame. A key feature of our
discussion is the notion of the temperature in the blackbody rest frame: the
frame in which the energy-momentum tensor is diagonal and the spectral
distribution is isotropic. This temperature is fixed, does not change under
Lorentz transformation. In this way we avoid the confused and confusing
question of how temperature transforms.\cite{nakamura} An important result
is that at zero temperature the spectral distribution is invariant under
Lorentz transformation. This is not an entirely new result, it has long been
recognized that in quantum electrodynamics the vacuum state is invariant,
but our derivation is explicit. Finally, at finite temperature the result
agrees with that obtained nearly half a century ago in connection with the
problem of detection of the earth's motion through the 2.7${{}^\circ}$
K cosmic radiation.\cite{heer68} However, as discussed in detail in \cite{nakamura}, there is a wide variety of opinion among the different authors in \cite{heer68}, in addition to many previous authors, as to how $T$ transforms, if at all.  Thus, our resolution of this question is an important part of our presentation.

We begin with the familiar expressions of quantum electrodynamics for the
free electric and magnetic field operators: \cite{sakurai67} 
\begin{eqnarray}
\mathbf{E}\left( \mathbf{r},t\right)  &=&\sum_{\mathbf{k},\alpha }\sqrt{
\frac{2\pi \hbar \omega }{V}}\left( ia_{\mathbf{k},\alpha }\hat{\mathbf{e}}
_{\alpha }e^{i(\mathbf{k}\cdot \mathbf{r}-\omega t)}-ia_{\mathbf{k},\alpha
}^{\dag }\hat{\mathbf{e}}_{\alpha }^{\ast }e^{-i(\mathbf{k}\cdot \mathbf{r}
-\omega t)}\right) ,  \notag \\
\mathbf{B}\left( \mathbf{r},t\right)  &=&\sum_{\mathbf{k},\alpha }\sqrt{
\frac{2\pi \hbar \omega }{V}}\left( ia_{\mathbf{k},\alpha }\mathbf{\hat{k}}
\times \hat{\mathbf{e}}_{\alpha }e^{i(\mathbf{k}\cdot \mathbf{r}-\omega
t)}-ia_{\mathbf{k},\alpha }^{\dag }\mathbf{\hat{k}}\times \hat{\mathbf{e}}
_{\alpha }^{\ast }e^{-i(\mathbf{k}\cdot \mathbf{r}-\omega t)}\right) ,
\label{qf1}
\end{eqnarray}
where $a_{\mathbf{k},\alpha }$ and $a_{\mathbf{k},\alpha }^{\dag }$ are the
usual lowering and raising operators for the field oscillators. Here we
should keep in mind that $\omega =ck$. The expressions (\ref{qf1}) were
obtained by quantizing in a box of volume $V$. We have in mind that the box
is at rest with walls at temperature $T$. The radiation within comes to
equilibrium at that temperature and we obtain the familiar result:\cite
{gerry}
\begin{equation}
\left\langle a_{\mathbf{k},\alpha }a_{\mathbf{k}^{\prime },\alpha ^{\prime
}}^{\dag }+a_{\mathbf{k}^{\prime },\alpha ^{\prime }}^{\dag }a_{\mathbf{k}
,\alpha }\right\rangle =\coth \left( \frac{\hbar \omega }{2kT}\right) \delta
_{\mathbf{k}\mathnormal{,}\mathbf{k}^{\prime }}\delta _{\alpha ,\alpha
^{\prime }},  \label{qf2}
\end{equation}
where the braces $\left\langle \cdots \right\rangle $ indicate the thermal
equilibrium expectation value. We shall need the correlation functions of
the field fluctuations: 
\begin{eqnarray}
C_{jm}^{(\text{\textnormal{el-el}})}\left( \mathbf{r}-\mathbf{r}^{\prime
},t-t^{\prime }\right)  &=&\frac{1}{2}\left\langle E_{j}\left( \mathbf{r}
,t\right) E_{m}\left( \mathbf{r}^{\prime },t^{\prime }\right) +E_{m}\left( 
\mathbf{r}^{\prime },t^{\prime }\right) E_{j}\left( \mathbf{r},t\right)
\right\rangle ,  \notag \\
C_{jm}^{(\text{\textnormal{mag-mag}})}\left( \mathbf{r}-\mathbf{r}^{\prime
},t-t^{\prime }\right)  &=&\frac{1}{2}\left\langle B_{j}\left( \mathbf{r}
,t\right) B_{m}\left( \mathbf{r}^{\prime },t^{\prime }\right) +B_{m}\left( 
\mathbf{r}^{\prime },t^{\prime }\right) B_{j}\left( \mathbf{r},t\right)
\right\rangle ,  \notag \\
C_{jm}^{(\text{\textnormal{el-mag}})}\left( \mathbf{r}-\mathbf{r}^{\prime
},t-t^{\prime }\right)  &=&\frac{1}{2}\left\langle E_{j}\left( \mathbf{r}
,t\right) B_{m}\left( \mathbf{r}^{\prime },t^{\prime }\right) +B_{m}\left( 
\mathbf{r}^{\prime },t^{\prime }\right) E_{j}\left( \mathbf{r},t\right)
\right\rangle .  \label{qf3}
\end{eqnarray}
With the expressions (\ref{qf1}) for the fields and (\ref{qf2}) for the
thermal expectation of the operators we use the prescription $
\sum_{k}\rightarrow \frac{V}{\left( 2\pi \right) ^{3}}\int d\mathbf{k}$ to
form the limit $V\rightarrow \infty $ and obtain the explicit results: 
\begin{eqnarray}
C_{jm}^{\mathnormal{(}\text{\textnormal{el-el}}\mathnormal{)}}\left( \mathbf{
r},t\right)  &=&C_{jm}^{\mathnormal{(}\text{\textnormal{mag-mag}}\mathnormal{
)}}\left( \mathbf{r},t\right)   \notag \\
&=&\frac{\hbar }{(2\pi )^{2}}\int d\mathbf{k}\omega \coth \frac{\hbar \omega 
}{2kT}\left( \delta _{jm}-\mathbf{\hat{k}}_{j}\mathbf{\hat{k}}_{m}\right)
\cos \left( \mathbf{k}\cdot \mathbf{r}-\omega t\right) ,  \notag \\
C_{jm}^{\mathnormal{(}\text{\textnormal{el-mag}}\mathnormal{)}}\left( 
\mathbf{r},t\right)  &=&\frac{\hbar }{(2\pi )^{2}}\int d\mathbf{k}\omega
\coth \frac{\hbar \omega }{2kT}e_{jml}\mathbf{\hat{k}}_{l}\cos \left( 
\mathbf{k}\cdot \mathbf{r}-\omega t\right) .  \label{qf4}
\end{eqnarray}

As a first application of these results, we calculate the spectral
distribution of blackbody radiation. The energy density of the
electromagnetic field in thermal equilibrium is given by 
\begin{equation}
W=\frac{\left\langle E^{2}\right\rangle +\left\langle B^{2}\right\rangle }{
8\pi }=\frac{C_{jj}^{(\text{\textnormal{el-el})}}(0,0)}{4\pi }.  \label{qf5}
\end{equation}
We can write 
\begin{equation}
W=\int_{0}^{\infty }d\omega \int d\Omega \rho \left( \omega ,\mathbf{\hat{k}}
\right) ,  \label{qf6}
\end{equation}
where $\rho \left( \omega ,\mathbf{\hat{k}}\right) d\omega d\Omega $ is the
energy density of radiation with frequency in the interval $d\omega $ and
propagating in solid angle $d\Omega $ about the direction $\mathbf{\hat{k}}$
. With the explicit results (\ref{qf4}) we find that the spectral
distribution in the rest frame is given by 
\begin{equation}
\rho \left( \omega ,\mathbf{\hat{k}}\right) =\frac{\hbar }{(2\pi c)^{3}}
\omega ^{3}\coth \frac{\hbar \omega }{2kT}.  \label{qf7}
\end{equation}
Of course, this is independent of the direction of propagation and, except
for a factor of $4\pi $, is just the Planck spectrum with the inclusion of
the zero point fluctuations.

We now consider the question of how this spectral distribution transforms
under a Lorentz transformation. Here we should remind ourselves that the
above discussion is understood to be in the rest frame of the blackbody
radiation: the frame in which the energy-momentum tensor is diagonal and the
spectral distribution is isotropic. The temperature $T$ is the temperature
in this rest frame and \emph{does not transform}. We begin with the
well-known expressions for the Lorentz transformation of the fields from a
frame at rest to a frame moving with velocity $\mathbf{v}$: \cite{jackson98} 
\begin{eqnarray}
\mathbf{E}^{\prime }\left( \mathbf{r}^{\prime },t^{\prime }\right)  &=&\hat{
\mathbf{v}}\cdot \mathbf{E}\left( \mathbf{r},t\right) \hat{\mathbf{v}}
+\gamma \left[ \mathbf{E}\left( \mathbf{r},t\right) -\hat{\mathbf{v}}\cdot 
\mathbf{E}\left( \mathbf{r},t\right) \hat{\mathbf{v}}+\frac{\mathbf{v}}{c}
\times \mathbf{B}\left( \mathbf{r},t\right) \right] ,  \notag \\
\mathbf{B}^{\prime }\left( \mathbf{r}^{\prime },t^{\prime }\right)  &=&\hat{
\mathbf{v}}\cdot \mathbf{B}\left( \mathbf{r},t\right) \hat{\mathbf{v}}
+\gamma \left[ \mathbf{B}\left( \mathbf{r},t\right) -\hat{\mathbf{v}}\cdot 
\mathbf{B}\left( \mathbf{r},t\right) \hat{\mathbf{v}}-\frac{\mathbf{v}}{c}
\times \mathbf{E}\left( \mathbf{r},t\right) \right] ,  \label{qf8}
\end{eqnarray}
where, as usual, $\gamma =1/\sqrt{1-v^{2}/c^{2}}$. With this we find for the
energy density in the moving frame:
\begin{eqnarray}
W^{\prime } &=&\frac{\left\langle E^{\prime 2}\right\rangle +\left\langle
B^{\prime 2}\right\rangle }{8\pi }  \notag \\
&=&\frac{1}{4\pi }\left\{ C_{jj}^{(\text{\textnormal{el-el}})}(0,0)+2(\gamma
^{2}-1)\left[ C_{jj}^{\left( \text{\textnormal{el-el}}\right) }(0,0)-\hat{
\mathnormal{v}}_{j}\hat{\mathnormal{v}}_{m}C_{jm}^{\left( \text{\textnormal{
\ el-el}}\right) }(0,0)\right] \right.   \notag \\
&&\left. +2\gamma ^{2}\frac{v_{l}}{c}e_{ljm}C_{jm}^{\left( \text{\textnormal{
\ el-mag}}\right) }(0,0)\right\} .  \label{qf9}
\end{eqnarray}
Using the expressions (\ref{qf4}) for the correlation functions together
with the expression (\ref{qf7}) for the spectral distribution, we can write
this in the form: 
\begin{equation}
\int_{0}^{\infty }d\omega ^{\prime }\int d\Omega ^{\prime }\rho ^{\prime
}\left( \omega ^{\prime },\hat{\mathnormal{k}}^{\prime }\right)
=\int_{0}^{\infty }d\omega \int d\Omega \gamma ^{2}\left( 1-\mathbf{\hat{k}}
\cdot \frac{\mathbf{v}}{c}\right) ^{2}\rho \left( \omega ,\mathbf{\hat{k}}
\right) .  \label{qf10}
\end{equation}
That is, 
\begin{equation}
\rho ^{\prime }\left( \omega ^{\prime },\mathbf{\hat{k}}^{\prime }\right)
=\gamma ^{2}\left( 1-\mathbf{\hat{k}}\cdot \frac{\mathbf{v}}{c}\right)
^{2}\rho \left( \omega ,\mathbf{\hat{k}}\right) \frac{d\omega }{d\omega
^{\prime }}\frac{d\Omega }{d\Omega ^{\prime }}.  \label{qf11}
\end{equation}

To proceed we introduce the Lorentz transformation of the propagation
vector: 
\begin{eqnarray}
\omega ^{\prime } &=&\gamma \left( \omega -\mathbf{k}\cdot \mathbf{v}\right) 
\notag \\
\mathbf{k}^{\prime } &=&\mathbf{k}-\hat{\mathbf{v}}\cdot \mathbf{k}\hat{
\mathbf{v}}+\gamma \left( \hat{\mathbf{v}}\cdot \mathbf{k}\hat{\mathbf{v}}
-\omega \frac{\mathbf{v}}{c^{2}}\right) .  \label{qf12}
\end{eqnarray}
For photons $\omega =ck$, which implies $\omega ^{\prime }=ck^{\prime }$. In
this case we get from the first Lorentz transform equation the formula for
the Doppler shift: 
\begin{equation}
\omega ^{\prime }=\gamma \left( 1-\mathbf{\hat{k}}\cdot \frac{\mathbf{v}}{c}
\right) \omega   \label{qf13}
\end{equation}
and from the second the aberration formula: 
\begin{equation}
\mathbf{\hat{k}}^{\prime }\cdot \hat{\mathbf{v}}=\frac{\mathbf{\hat{k}}\cdot 
\hat{\mathbf{v}}-\frac{v}{c}}{1-\mathbf{\hat{k}}\cdot \frac{\mathbf{v}}{c}}.
\label{qf14}
\end{equation}
It will be useful to have the inverse of these formulas, obtained by solving
the aberration formula for $\mathbf{\hat{k}}\cdot \mathbf{\hat{v}}$ and then
putting the result in the Doppler shift formula. The result is
\begin{equation}
\mathbf{\hat{k}\cdot \hat{v}=}\frac{\mathbf{\hat{k}}^{\prime }\mathbf{\cdot 
\hat{v}}+\frac{v}{c}}{1+\mathbf{\hat{k}}^{\prime }\mathbf{\cdot }\frac{
\mathbf{v}}{c}},\quad \omega =\gamma \left( 1+\mathbf{\hat{k}}^{\prime }
\mathbf{\cdot }\frac{\mathbf{v}}{c}\right) \omega ^{\prime }  \label{qf15}
\end{equation}
From these formulas, we find
\begin{eqnarray}
d\omega  &=&\gamma \left( 1+\mathbf{\hat{k}}^{\prime }\mathbf{\cdot }\frac{
\mathbf{v}}{c}\right) d\omega ^{\prime },  \notag \\
d\Omega  &=&\frac{d\mathbf{\hat{k}\cdot \hat{v}}}{d\mathbf{\hat{k}}^{\prime }
\mathbf{\cdot \hat{v}}}d\Omega ^{\prime }=\frac{d\Omega ^{\prime }.}{\gamma
^{2}\left( 1+\mathbf{\hat{k}}^{\prime }\mathbf{\cdot }\frac{\mathbf{v}}{c}
\right) ^{2}}.  \label{qf16}
\end{eqnarray}
With these results in the right hand side of the identity (\ref{qf11}) we
find that the spectral distribution in the moving frame is given by
\begin{equation}
\rho ^{\prime }\left( \omega ^{\prime },\mathbf{\hat{k}}^{\prime }\right) =
\frac{\rho \left( \gamma \left( 1+\mathbf{\hat{k}}^{\prime }\mathbf{\cdot }
\frac{\mathbf{v}}{c}\right) \omega ^{\prime },\mathbf{\hat{k}}\right) }{
\gamma ^{3}\left( 1+\mathbf{\hat{k}}^{\prime }\mathbf{\cdot \frac{\mathbf{v}
}{c}}\right) ^{3}},  \label{qf17}
\end{equation}
Finally, with the expression (\ref{qf7}) for the spectral distribution in
the rest frame, we find that in the moving frame it takes the explicit form: 
\begin{equation}
\rho ^{\prime }\left( \omega ^{\prime },\mathbf{\hat{k}}^{\prime }\right)
=\hbar \left( \frac{\omega ^{\prime }}{2\pi c}\right) ^{3}\coth \left( \frac{
\hbar \gamma \left( 1+\mathbf{\hat{k}}^{\prime }\cdot \frac{\mathbf{v}}{c}
\right) \omega ^{\prime }}{2kT}\right) ,  \label{qf18}
\end{equation}

The expression (\ref{qf18}) is our key result. First, we note that at zero
temperature this spectral distribution in the moving frame is exactly of the
form of that in the rest frame. That is, the spectral distribution at zero
temperature is invariant under Lorentz transformations. Moreover, at finite
temperature our result is exactly of the form long known in discussions of
motion through the 2.7${{}^\circ}$ K cosmic radiation.\cite{heer68} However, our derivation has made it clear
that $T$ is the invariant temperature in the blackbody rest frame. There has
therefore been no need to get into the question of how temperature
transforms under Lorentz transformations: our $T$ is the temperature in the
rest frame and does not change.

This work was partially supported by the National Science Foundation under
Grant No. ECCS-1125675.


\begin{thebibliography}{9}
\bibitem{nakamura} T. K. Nakamura, \emph{Europhys. Lett.} \textbf{88}, 20004
(2009), Z. C. Wu, Europhys. Lett. \textbf{88}, 20005 (2009).

\bibitem{heer68} C. V. Heer and R. H. Cole, \emph{Phys. Rev.} \textbf{174}, 1611
(1968); P. J. E. Peebles and D. T. Wilkinson, \emph{Phys. Rev.} \textbf{174}, 2168 (1968); R. N. Bracewell and E. K. Conklin, \emph{Nature} \textbf{219}, 1343 (1968); G. R. Henry, R. B. Feduniak, J. E. Silver and M. A. Peterson, \emph{Phys. Rev.} \textbf{176}, 1451 (1968).

\bibitem{sakurai67} J. J. Sakurai, "Advanced quantum mechanics"
(Addison-Wesley, Reading, Massachusetts, 1967) p. 32.

\bibitem{gerry} C. C. Gerry and P. L. Knight, "Introductory Quantum Optics,"
p. 22, has in essence this result.

\bibitem{jackson98} J. D. Jackson, \emph{Classical Electrodynamics}, Second
Edition (Wiley, New York, 1998). p. 552.
\end{thebibliography}
\end{document}